\documentclass[12pt]{iopart}

\usepackage{graphicx, caption}
\usepackage{color}

\usepackage{hyperref}
\hypersetup{ 
	colorlinks=true,
	urlcolor=blue,citecolor=blue}

\begin{document}
\title[Ultrafast Spectroscopy of semiconductor and plasmonic nanostructures]{Ultrafast optical spectroscopy of semiconducting and plasmonic nanostructures and their hybrids.}

\author{Daniele Catone$^1$, Lorenzo Di Mario$^1$\footnote{present address: University of Groningen, Groningen, The Netherlands}, Faustino Martelli$^2$, Patrick O{'}Keeffe$^{3,*}$,  Alessandra Paladini$^3$, Jacopo Stefano Pelli Cresi$^3$\footnote{present address:  Sincrotrone Trieste S.C.p.A., Trieste, Italy}, Aswathi K. Sivan$^2$, Lin Tian$^2$\footnote{present address: Institute for Quantum Computing and Department of Electrical and Computer Engineering, University of Waterloo, ON N2L 3G1, Waterloo, Canada}, Francesco Toschi$^3$, Stefano Turchini$^1$}

\address{$^1$ Istituto di Struttura della Materia-CNR (ISM-CNR), Division of Ultrafast Processes in Materials (FLASHit), 100 Via del Fosso del Cavaliere, 00133 Rome, Italy.}
\address{$^2$ CNR-IMM, Area della Ricerca di Roma Tor Vergata, 100 Via del Fosso del Cavaliere, 00133 Rome, Italy.}
\address{$^3$ Istituto di Struttura della Materia-CNR (ISM-CNR), Division of Ultrafast Processes in Materials (FLASHit), 00015 Monterotondo Scalo, Italy.}
\ead{$^*$patrick.okeeffe@ism.cnr.it}

\begin{abstract}
The knowledge of the carrier dynamics in nanostructures is of fundamental importance for the development of (opto)electronic devices. This is true for semiconducting nanostructures as well as for plasmonic nanoparticles (NPs). Indeed, improvement of photocatalytic efficiencies by combining semiconductor and plasmonic nanostructures is one of the reasons why their ultrafast dynamics are intensively studied. In this work, we will review our activity on ultrafast spectroscopy in nanostructures carried out in the recently established EuroFEL Support Laboratory. We have investigated the dynamical plasmonic responses of metal NPs both in solution and in 2D and 3D arrays on surfaces, with particular attention being paid to the effects of the nanoparticle shape and to the conversion of absorbed light into heat on a nano-localized scale. We will summarize the results obtained on the carrier dynamics in nanostructured perovskites with emphasis on the hot-carrier dynamics
and in semiconductor nanosystems such as ZnSe and Si nanowires, with particular attention to the band-gap bleaching dynamics. Subsequently, the study of semiconductor-metal NP hybrids, such as CeO$_2$-Ag NPs, ZnSe-Ag NPs and ZnSe-Au NPs, allows the discussion of interaction mechanisms such as charge carrier transfer and F{\"o}rster interaction. Finally, we assess an alternative method for the sensitization of wide band gap semiconductors to visible light by discussing the relationship between the carrier dynamics of TiO$_2$ NPs and V-doped TiO$_2$ NPs and their catalytic properties.

\end{abstract}
%\keywords{magnetic moment, solar neutrinos, astrophysics}
%\submitto{\jpg}
\maketitle

\section{Introduction}

The study of hybrid nanostructured semiconductor and plasmonic materials has undergone an explosion in the 
recent years largely thanks to the hope of improving the performances of semiconductor devices and materials
by taking advantage of the strong light/matter interaction of the plasmonic nanomaterials~\cite{Clavero14,Liu17,Zhang18}. A particularly fertile
area of research is the sensitization of wide band gap semiconductors to visible light by this method. For example, transition metal
oxide catalysts such as TiO$_2$ and ZnO have band gaps larger than 3 eV which means that they can only absorb light in the UV 
region of the spectrum which contains $<$ 5 \% of the solar energy. This means that the 
efficiency of photocatalysis by such materials averaged over the entire spectrum is intrinsically low. This limitation
has largely blocked the widespread use of these materials for green photocatalysis methods such as
water splitting, CO$_2$ reduction and water disinfection~\cite{Valenti16,Wu18,Galan20}. Furthermore, in recent years there is a growing
interest in employing plasmonic hot electrons in hybrid devices for sensing, photodetection and solar energy applications~\cite{Tang20}.  

The physical basis for the strong light matter interaction of plasmonic nanoparticles (NPs) is that at certain (resonant) wavelengths the conduction electrons in the nanostructure start to oscillate in resonance with the frequency of the light. Thanks to this localized surface plasmon resonance (LSPR) an extremely intense local concentration of the electric field of the light into ``hot spots” is achieved. The order of magnitude of the field enhancement is expressed by the so-called quality factor Q=$\omega$/2$\gamma$ ~\cite{Stockman11}, where $\omega$ is the angular frequency of the plasmon resonance and $\gamma$ is the total decay rate. It should be noted that the relaxation dynamics plays a fundamental role in engineering plasmonic devices. Furthermore, these objects exhibit extremely large absorption and scattering cross sections which can be up 5 orders of magnitude larger than common organic dye molecules for example~\cite{Jain06}. 

Incorporating plasmonic materials into semiconductor nanostructures aims to take advantage of these large absorption/scattering cross sections by any one of 
a number of different mechanisms such as: i) light trapping, ii) near-field electromagnetic field concentration iii) resonant energy transfer, iv) indirect hot electron injection and v) direct decay of a hybridized plasmonic/semiconductor 
state into the semiconductor~\cite{Wu18}. Light trapping can occur by multiple scattering events within a 3D plasmonic nanostructure thanks to the large scattering cross-sections. This leads to an increased absorption path and thus to higher extinction efficiency. Near-field electromagnetic field concentration can be viewed as light concentration due to the effect of the plasmon oscillation. The result is increased absorption near the surface of the plasmonic material and can be taken advantage of by physically putting the plasmonic and semiconductor materials in contact. The resonant energy transfer can be mediated by dipole-dipole coupling leading to the transfer of energy between these dipoles. This interaction, which has long been known to take place between donor and acceptor molecules, is known as F{\"o}rster Resonant Energy Transfer (FRET) and can also take place in hybrid plasmonic/semiconductor nanostructures~\cite{Wu15,Li15}. All of these mechanisms require a spectral overlap of the absorption in the two materials. On the other hand the last two mechanisms do not require such overlap and the semiconductor can be excited by decay of the plasmonic state. In the case of the indirect hot electron injection the decay of the plasmon produces a non-thermal electron distribution in the plasmonic structure. This can be followed by injection of electrons over a Schottky barrier formed at the plasmonic material/semiconductor interface and is also referred to as an internal photoemission process. The channels for the formation of hot electrons are~\cite{Khurgin19}: free carrier absorption arising from phonon and defect scattering, absorption assisted by electron-electron scattering, absorption via Landau damping. In the latter, hot electrons are generated in the near surface region by a surface collisions assisted mechanism and they have a good average directionality with respect to the electric field. Landau damping is considered the more promising channel to overcome the Schottky barrier and to inject hot electrons in the semiconductor. This kind of damping is particularly efficient in nanoparticles whose size is less than the mean electron path. In the direct decay mechanism the plasmon can dephase into empty hybridized orbitals formed at the interface 
between the plasmonic and semiconductor materials~\cite{Tan17}. If these orbitals have significant semiconductor character the result is direct
formation of hot electrons in the semiconductor as a result of decay of the plasmon.   

Which one of these mechanisms, if any, dominates depends on a large number of factors including: i) alignment of the bands of materials, 
ii) excitation wavelength, iii) morphology of the materials, iv) the presence and location of hotspots, v) the relative strength 
of absorption and scattering of the plasmonic material vi) structure of the interface, and vii) the presence and type of defects in the semiconductor material, to name just a few. 

In our work we study these interactions by monitoring the temporal evolution of the optical response of
the materials following photoexcitation. To be able to understand and correctly interpret the response
of the hybrid systems it is clearly necessary to first understand the response of the individual plasmonic and 
semiconductor materials. The ultrafast processes in the two materials are quite different so we will discuss
them separately.  

The optical response of metal nanoparticles have been extensively studied over the years~\cite{Hartland11}. Following the photoexcitation of the LSPR of a plasmonic nanostructure, the energy absorbed by the
nanoparticle is transferred from the oscillating electrons of the plasmon to conduction electrons mainly by Landau damping resulting in 
a non-thermal energy distribution of the electrons above and below the Fermi level on timescale of tens of femtoseconds. 
The resulting energetic electrons then undergo a thermalization process through electron-electron scattering to produce a Fermi-Dirac
energy distribution characterized by a temperature of hundreds to thousands of Kelvin depending on the energy and intensity of
the exciting light. This process generally takes place within hundreds of femtoseconds. 
Subsequently, this hot electron distribution undergoes cooling mainly by electron-phonon coupling
in which energy is transferred to the lattice of the plasmonic nanostructure within several picoseconds. 
Finally, the plasmonic nanostructure exchanges energy with the surrounding materials through phonon-phonon 
coupling typically on the scale of hundreds of picoseconds. The precise timings of these processes are strongly dependent
on the quantity of energy deposited in the nanostructure during the laser pulse, with the processes generally taking longer for higher intensity excitation. All of these processes can be monitored by measuring the optical response of the 
plasmonic nanostructures in time as will be discussed below. In addition to the above processes, competing
phenomena can take place when the intensity of the exciting light is raised above a threshold which can result
in modification of the morphology of the nanostructures or non-linear processes.          

In a nanostructured semiconductor, absorption of light with energy above the band gap of the material
leads to promotion of an electron from the valence band to the conduction band, resulting in the formation
of an electron-hole pair. In the case in which the energy of the photon is significantly above the 
band gap, a very energetic non-thermal distribution of charge carriers is produced. This distribution rapidly
thermalizes through carrier-carrier scattering on the 100s of femtoseconds timescale thus allowing
the energy distribution to be described by a Fermi-Dirac distribution with a well defined temperature. 
Subsequently, energy relaxation to the band edges can take place by emission of optical phonons on the picosecond
timescale. The signals can also decay via three body Auger-type processes, band-to-band 
recombination and trap-assisted (Schokley-Read-Hall) recombination, each of which have characteristic lifetimes. The time-dependent carrier density can be monitored down 
to the femtosecond timescale by following the bleaching of the band edge absorption in
transmission experiments.

These dynamics are strongly dependent on the excitation intensity and thus the initial carrier density. For example, 
many body effects such as band gap renormalization can lead to a time-dependent red-shift of the band edge. Other 
intensity dependent dynamics include band filling which leads to a time-dependent blueshift of
the absorption edge and non-equilibrium phonon populations which can slow the relaxation of the carriers (hot-phonon
bottleneck). Clearly, also the band-to-band and Auger recombination processes depend on the carrier density.
Furthermore, the dynamics can depend strongly on the defect density of the semiconductor as trapping and trap-assisted decay 
can modify the lifetimes of the carriers. An excellent review on the carrier and phonon dynamics in semiconductors can be found in~\cite{Othonos98}.

 As can be seen from the above discussion, the dynamics of both the plasmonic nanoparticles and the semiconductor nanostructures
 on the femtosecond timescales are both complicated and intensity dependent and can be influenced by numerous factors. To gain some insight into the extraction of reliable information on these dynamics from optical pump-probe methods we structure
 this article in the following manner: first, Transient Absorption (TA) spectroscopy (the main experimental optical technique used in our studies) is briefly described, we then summarize some of our
 measurements on plasmonic metal nanoparticles~\cite{Fratoddi18,Catone18,Magnozzi19,Dimario18}, this is followed by a discussion of the dynamics extracted from our optical measurements on a range of
 nanostructured semiconductors~\cite{Tian16,Tian19,Okeeffe19}, we then  
 analyze the dynamics in hybrid systems~\cite{PelliCresi19,Sivan20} with the aim of achieving an understanding of the 
 parameters governing the modification of the dynamics due to interactions in these systems and, finally,
 we describe the carrier dynamics of TiO$_2$ NPs and V-doped TiO$_2$ NPs and how they affect their catalytic properties.

\section{Methods}
\subsection{Ultrafast transient absorbance spectroscopy}

Ultrafast TA spectroscopy pump-probe measurements were all performed using a femtosecond laser system consisting of an oscillator (20 fs pulse duration, 80 MHz repetition rate, 550 mW power centred at 800 nm) that seeds a regenerative amplifier thus producing 35 fs pulses at 1 KHz with 4 mJ peak power. The pump pulses are produced by using the output of an optical parametric amplifier in the 240 - 1600 nm wavelength range. The probe is a white light supercontinuum beam (250 - 1550 nm) that is generated in a commercial TA spectrometer (FemtoFrame II, IB Photonics) using three different setups depending on the energy range to be covered (UV, visible, and IR): the UV probe (240 - 370 nm) is generated by focusing a 400 nm laser beam into a rotating CaF$_2$ crystal; the visible probe (360 - 780 nm) is generated by focusing an 800 nm pulse into a rotating CaF$_2$ crystal; the IR probe (850 - 1550 nm) is generated by focusing an 800 nm pulse into a YAG crystal.
The TA spectroscopy experiments are performed using a pump and a probe beam focused on the sample (pump diameter = 300 $\mu$m; probe diameter = 150 $\mu$m) and modifying the delay time between the two by changing the optical path of the probe. In the case of the UV and visible probe, the beam is split into two with a half collected by the first spectrometer and used as reference to account for pulse to pulse fluctuations in the white light generation. The remaining half probe beam is collected by the second spectrometer after being transmitted or reflected by the sample. In IR probe a single spectrometer is used. The differences in absorbance (reflectance) of the sample when excited by the pump and when unperturbed, i.e. the transient absorbance $\Delta$A (transient reflectance $\Delta$R), are measured as a function of the pump-probe delay time. These setups allow both liquid and solid samples to be investigated with an instrument response function of about 50 fs. Moreover, the solid samples can be measured in air at room temperature or under vacuum at 77K. More details can be found elsewhere~\cite{Okeeffe19,PelliCresi19}.

\section{Results and discussion}

\subsection{Ultrafast Optical Response of Nanostructured Plasmonic Materials}

\subsubsection{Gold Nanoparticles in Solution \\ }

\begin{figure}[htb]
	\captionsetup{singlelinecheck = false, font=footnotesize, labelsep=space}
	\centering
	\includegraphics[scale=0.5]{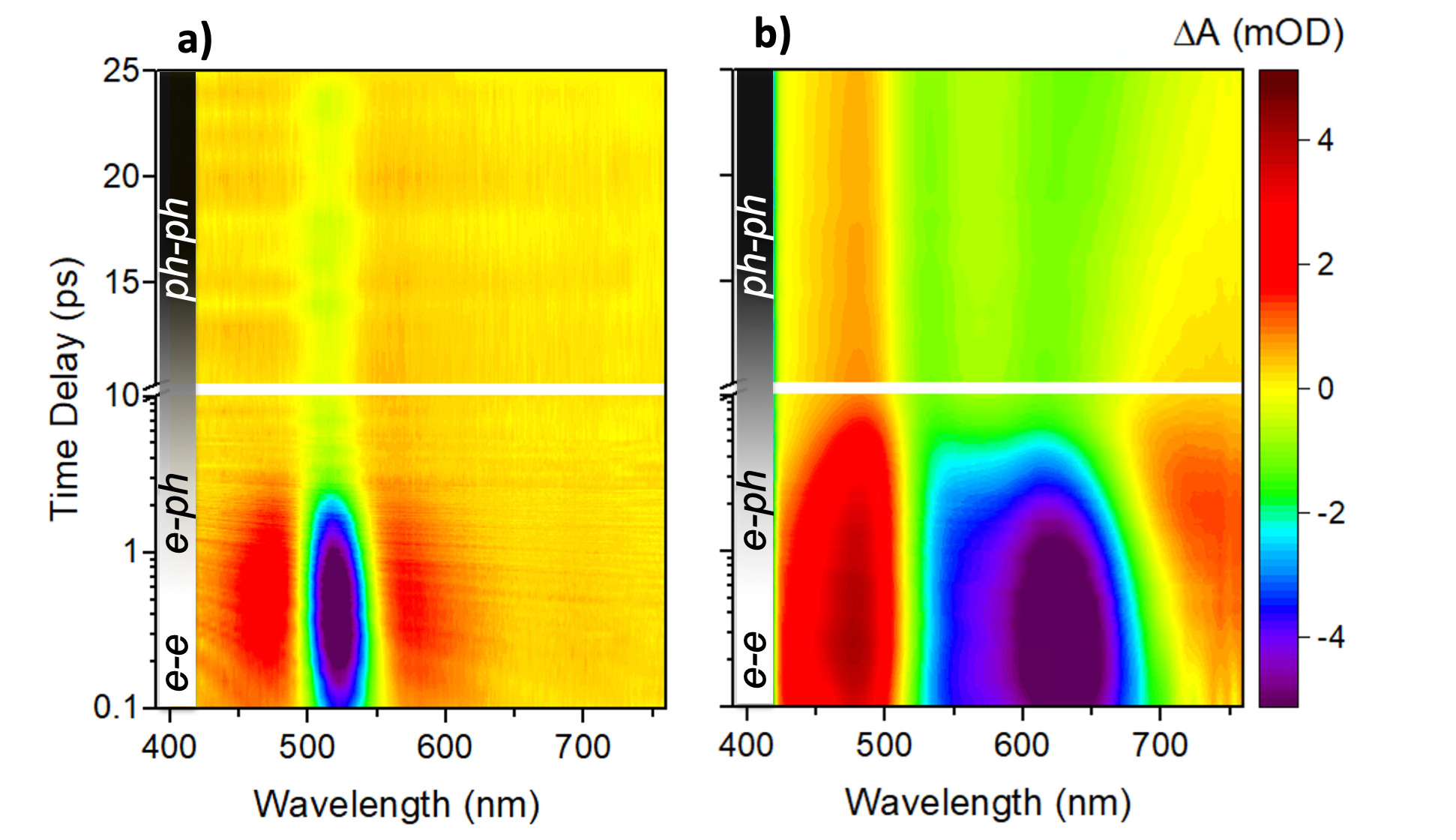}
	\caption{ Transient absorption map of a water solution of: a) 10 nm citrate capped gold spherical NPs excited with a 370 nm pump, b) 10 nm gold spherical NPs, functionalized by rhodamine B isothiocyanate (Au@RITC)~\cite{Fratoddi18} excited with a 400 nm pump. On the left of each map, a vertical stripe indicates the temporal range of the subsequent scattering (e-e, e-ph, ph-ph) events following photoexcitation of the nanostructure
}
	\label{NPs_sol}
\end{figure}

The optical response of plasmonic materials following photoexcitation can be best illustrated by pump-probe studies on gold spherical 
nanoparticles. Figure \ref{NPs_sol}a reports the false-color TA map of a water solution of 10 nm citrate capped Au NPs (0.05 mg/ml), when excited by a 
370 nm pump beam with a fluence of 150 $\mu$J/cm$^2$. The map is characterized by a strong negative feature around 520 nm and two positive wings on both 
sides, the one on the blue side being much more intense. These features are due to the alterations of the optical properties of the Au NPs upon excitation by the femtosecond pump beam. Following irradiation, the electrons heat up~\cite{Hartland11} causing a depletion and broadening of the plasmon resonance. As a result, a (negative) photobleaching (PB) of the probe signal and the appearance of two positive wings on either sides of the bleaching is observed in the TA map~\cite{Ahmadi96}.

During the first few hundreds of femtoseconds, the features of these signals moves to shorter wavelengths and increases until the nascent 
non-thermal electron distribution has thermalized to a hot distribution. After reaching their maximum values, the signals of both the 
bleaching and the positive wings decrease in a few picoseconds, mainly due to electron-phonon coupling~\cite{Hartland11}. Finally, a long timescale 
decay is observed, as the NP returns to its unperturbed state on a timescale of tens/hundreds of picoseconds through phonon-phonon scattering processes.

As is well known, the wavelength position and bandwidth of LSPRs depend on a number of factors, including the size, shape and composition of the 
plasmonic materials~\cite{Hartland04}. As visible in figure \ref{NPs_sol}a, spherical NPs exhibits only a single (transverse) LSPR due to their high-order symmetry, while 
NPs with different symmetries present more LSPR modes~\cite{El-Sayed01,Murphy05}. 

Interesting effects are also observed when NPs approach each other and interact. We studied 10 nm gold spherical NPs, functionalized by rhodamine B 
isothiocyanate (Au@RITC)~\cite{Fratoddi18}: by modulating the quantity of dye molecules on the surface of the nanoparticles, and in turn the 
dielectric constant of the material between the nanoparticles, it was possible to exploit the coupling of plasmonic resonances between closely 
spaced NPs~\cite{Kravets16,Kojima11}. Such a coupling can be described in terms of a delocalization of the plasmon resonance over a number of 
particles~\cite{Ghosh07,Jain10}, leading to the appearance of an additional extended LSPR. Figure \ref{NPs_sol}b shows the false-colour TA map of the Au@RITC 
water solution, when excited by a 400 nm pump beam with a fluence of 180 $\mu$J/cm$^2$. The bleaching of both the transverse and extended LSPRs are 
observable, around 530 nm and 630 nm, respectively. The broadness of the extended LSPR suggests the presence of many substructures~\cite{Fratoddi18}, such as chains of different length, bidimensional nanoaggregates, and so on.

\subsubsection{3D Arrays of Gold Nanoparticles deposited on SiO$_2$ Nanowires \\ }

We now discuss how the simple deposition of the plasmonic NPs on the surface of the semiconductor may change the properties of the metallic NPs themselves. This is particularly true when NPs are deposited on the sidewalls of nanowires (NWs). The reason for changing the plasmonic characteristics is that the NW sidewalls induce a NP shape that is not spherical or hemispherical, but ellipsoidal. The silica NW arrays have been fabricated via thermal oxidation of Si NWs grown on a quartz substrate. The silica NWs are an excellent support for metal NPs because they provide a large surface area to attach the NPs and, at the same time, a macroporous framework that ensures an efficient interaction with the environment. The arrays are also fully transparent in the visible to near-UV region of the spectrum, which helps the characterization of the NPs because allows the measurements in transmission. Additionally, the strong light scattering typical of the nanowire-based structures increases the light absorption of the NPs. Silica NW arrays have been decorated with both Au~\cite{Dimario18} and Ag NPs, which have been obtained by dewetting thin metallic films evaporated on top of the NWs~\cite{Convertino14}. 

\begin{figure}[htb]
	\captionsetup{singlelinecheck = false, font=footnotesize, labelsep=space}
	\centering
	\includegraphics[scale=1.0]{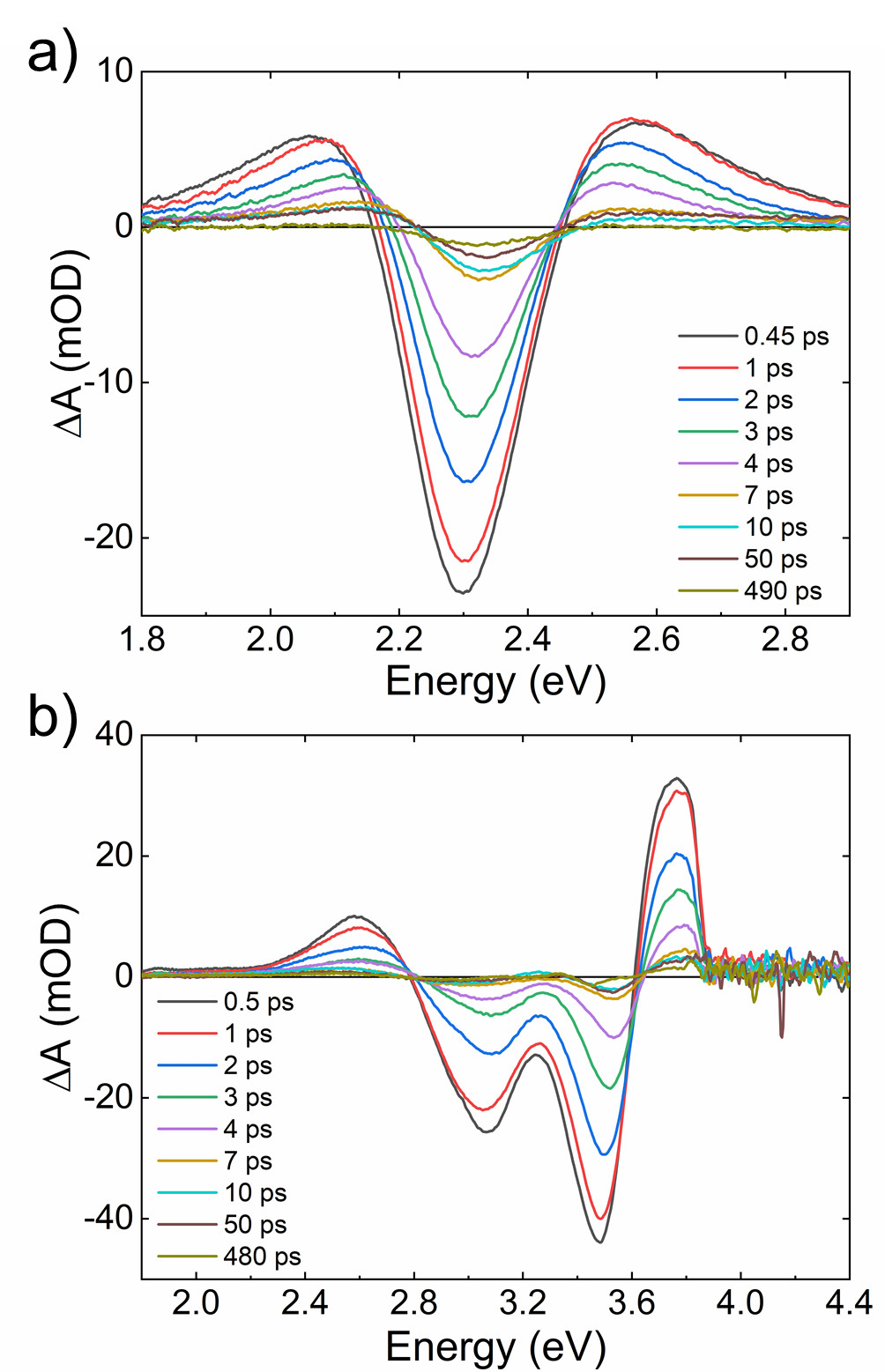}
	\caption{Transient absorbance spectra at different pump-probe delay time for metal NPs on silica NWs: a) Au NPs, b) Ag NPs. }
	\label{NPSiO2}
\end{figure}

\begin{figure}[htb]
	\captionsetup{singlelinecheck = false, font=footnotesize, labelsep=space}
	\centering
	\includegraphics[scale=1.0]{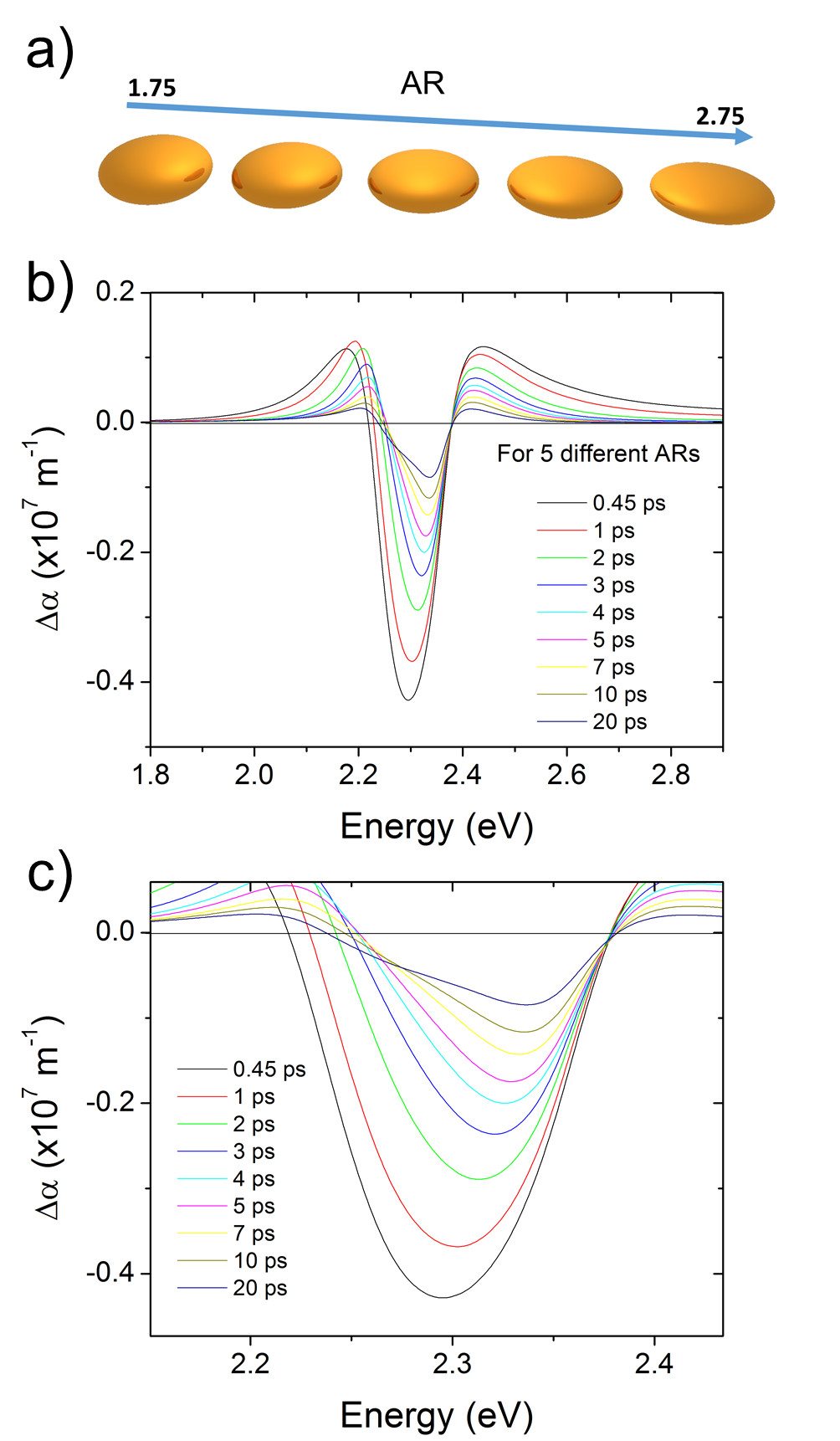}
	\caption{ a) Graphic representation of the five aspect ratios (1.75, 2.00, 2.25, 2.50, and 2.75) considered for the oblate Au NPs. b) Calculated spectral dependence of $\Delta$A for NPs with different ARs and their average value at 4 ps delay. c) Close-up of bleaching position, which shows the blueshift with increasing delay time. }
	\label{NPSiO2_theory}
\end{figure}

In figure \ref{NPSiO2} the spectral dependence of the transient absorbance at different pump-probe delay times is reported for both Au (figure \ref{NPSiO2}a) and Ag-decorated silica NWs (figure \ref{NPSiO2}b). As expected, the spectra show a decrease of the TA signal at the LSPR immediately after the pump excitation. For the Ag NPs both the dipole and quadrupole contribute to the plasmon resonance. All the probe energies show similar dynamics, indicating that a single process is dominant in this time range: the electron-phonon coupling. An interesting feature of our results is the blue-shift of the TA resonance bleaching for increasing delay time. This feature is unusual for homogeneous distributions of metallic NPs and we have attributed it to the shape distribution of the Au and Ag NPs resulting from the dewetting method used to fabricate them. Using a three-temperature model (3TM)~\cite{Dimario18} that includes the electron T$_e$(t) and lattice T$_l$(t) temperatures of the NPs and the temperature of the surrounding environment T$_{env}$(t), we have observed that to account for the observed shift we had to consider the NP shape and its distribution. In particular, for Au NPs, we have considered oblate NPs with different aspect ratio to be the closest picture to the real shape distribution obtained with our sample preparation. In figure \ref{NPSiO2_theory}a a graphic representation of the shape distribution used for the simulations can be found.

Thanks to our simulations (figure \ref{NPSiO2_theory}b) that account for the real shape distribution of our NPs, we have been able to assign several characteristics observed in the experimental transient optical response to the effect of the distribution of the NP shape. In particular, the profile of the two positive wings is strongly related to the shape and to the shape distribution of the NPs, as expected from the observation of quite different behaviors in spherical and rodlike NPs~\cite{Guillet09,Guillet09b,Baida11}. The figure \ref{NPSiO2_theory}b shows that at a given delay time the position of the absorption bleaching is different for different aspect ratios of the oblate. When the shape distribution is then considered in the simulation of the overall system, the result shows a blue shift of the bleaching position for increasing pump-probe delay times, as shown in figure \ref{NPSiO2_theory}c. The shape distribution, that has been accurately calculated for Au NPs~\cite{Dimario18}, is assumed to be the reason for the similar blue-shift of the LSPR bleaching observed for Ag NPs.

These observations have important consequences when examining the transient responses of mixed NP/semiconductor systems: ultrafast
changes in the spectral response of the LSPR bleaching may not be related to ultrafast
energy transfer processes between the materials but may simply be due to wide morphological distributions of NPs within the sample.\\

\subsubsection{Thermal Effects in the Optical Excitation of Plasmonic Materials  \\}

As can be seen from the above discussion most of the energy absorbed by plasmonic materials such as metal NPs gets
rapidly converted into heat. While it can be interesting to deposit heat in a controlled manner on the nanometer scale~\cite{Baffou13}, in mixed 
plasmonic/semiconductor systems aimed at transferring absorbed energy from the plasmonic material to the semiconductor it is an undesirable process. In a simple energy balance equation the energy absorbed by the plasmonic material but not converted into heat 
has been transferred to the surrounding material. The latter is the desired process in the above plasmonic/semiconductor system.
Therefore, in this section we discuss methods to measure the quantity of energy desposited as heat in metal NPs systems and how 
different types of exciting light deposit thermal energy in different ways.   

\begin{figure}[htb]
	\captionsetup{singlelinecheck = false, font=footnotesize, labelsep=space}
	\centering
	\includegraphics[scale=0.35]{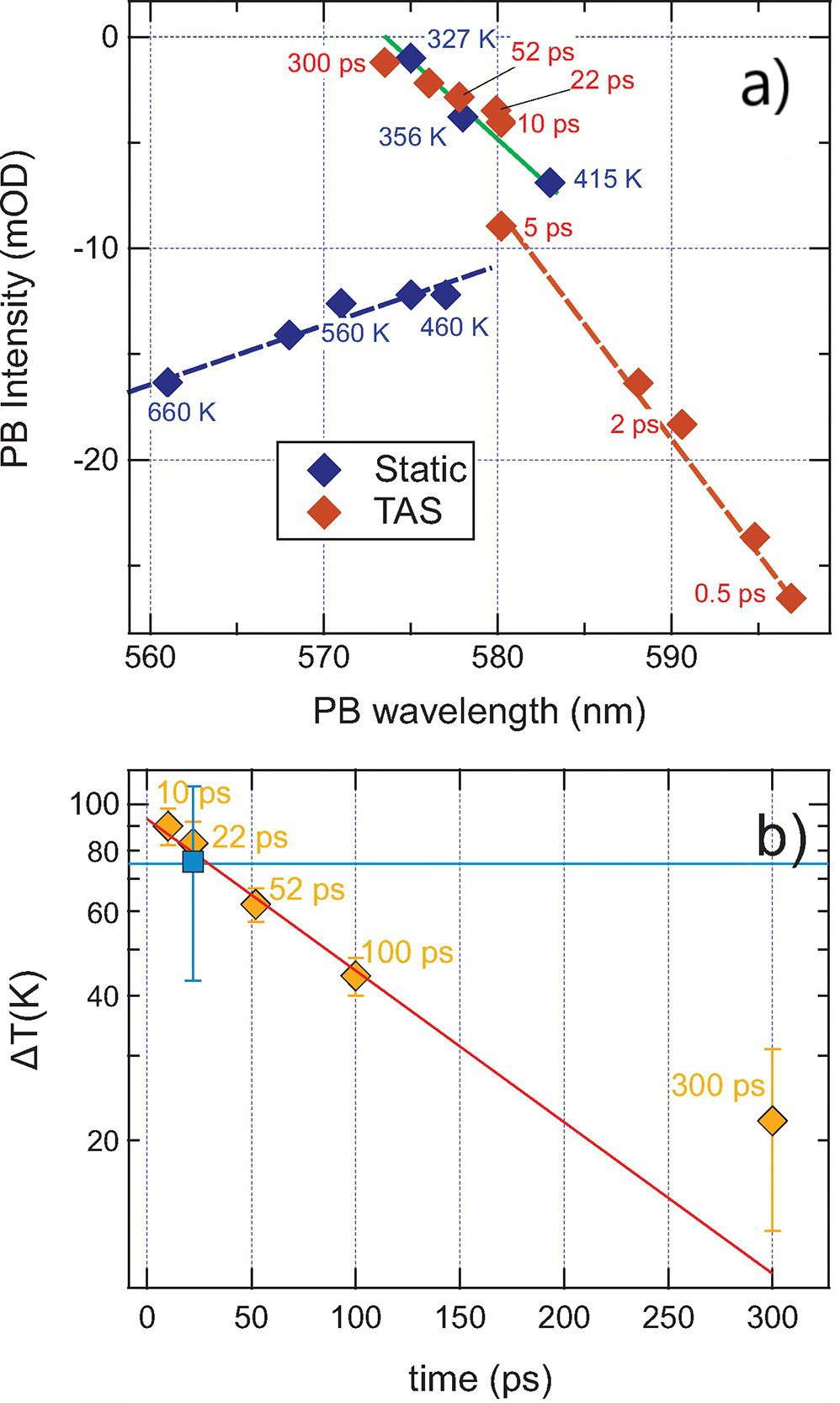}
	\caption{a) Plot of the photobleaching intensity against photobleaching wavelength extracted from the dynamics of the TA spectra (red markers) and those extracted from static T$_{bath}$-dependent data (blue markers). The values of $\tau$ and T$_{bath}$ of each point is indicated in the figure. The dashed lines act as a guide to the eye while the continuous green line is an interpolation of the static data in the 415-327 K T$_{bath}$ range. b) The NP temperature extracted from the analysis of the data in panel a) (orange symbols) plotted as as a function of time together with the estimated low-delay NP temperature (blue symbols). The red line shows an exponential fit to the orange points. Reprinted and adapted with permission from ~\cite{Ferrera20}. Copyright (2020) American Chemical Society.}
	\label{Bisio_Therm}
\end{figure}

With these concepts in mind we have developed a method to determine the time dependent temperature of the NPs in the relaxation processes after an ultrashort pulse excitation. The electronic temperature (T$_e$), lattice temperature (T$_l$), and environmental temperature (T$_{env}$) are the basic ingredients of theoretical models such as two and three temperature models~\cite{Zavelani15} such as that described above for the 3D arrays of AuNPs on SiO$_2$ nanowires~\cite{Dimario18}. Generally, the temperature of the NPs are extrapolated indirectly from experimental data by comparison of the pump-probe transient absorption to the results of these models. In our work~\cite{Ferrera20} a simple method to estimate the temporal evolution of the temperature has been proposed and demonstrated on self-organized 
2D arrays of ellipsoidal AuNPs deposited onto the surface of a LiF single crystal substrate~\cite{Anghinolfi12}. The result was achieved by comparing data obtained from the temporal evolution of the transient photobleaching peak and the static absorption spectra of the AuNP system immersed in a thermodynamic bath where the temperature (T$_{bath}$) is known. In the latter case by definition T$_e$=T$_l$=T$_{env}$=T$_{bath}$. Overlapping the two experimental datasets it was possible to determine a good match between the dynamic and static data associating a temperature with a delay time as shown in (figure \ref{Bisio_Therm} a). It should be noted that for times below 10 ps the static data do not match with dynamic ones, thus suggesting that it is the temporal lower limit of the method. The main reason for this is that at these delay times T$_e$ and T$_l$ are still strongly out of equilibrium 
because the system is still transferring the energy/heat from electrons to the lattice. By fitting the data from the longer times ($>$ 10ps) with a linear function and projecting this onto the data of the lower T$_{bath}$ static points it was possible to build an effective thermometric scale (figure \ref{Bisio_Therm}b). Then the experimental TA points can be projected onto the calibration curve estimating the associated temperature. This method is not limited to the system of AuNPs described above but it is applicable to all systems where it is possible measure the ultrafast optical response and steady state spectral response at controlled temperature.

Having a method to extract the temperature of the NPs will be of use when evaluating the amount of absorbed energy
which is converted into heat in hybrid plamonic/semiconductor systems as it may provide a tool to evaluate how much
energy is transferred from the NP to the semiconductor material in the first 100s of femtoseconds before the heating takes place.

\begin{figure}[htb]
	\captionsetup{singlelinecheck = false, font=footnotesize, labelsep=space}
	\centering
	\includegraphics[scale=0.35]{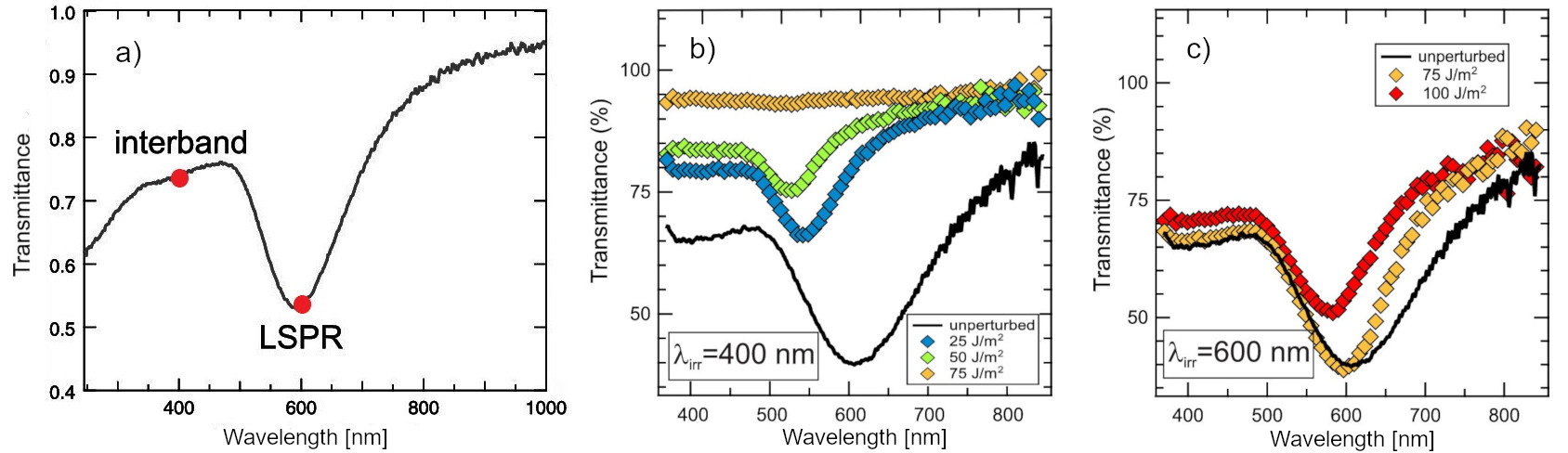}
	\caption{ a) Transmission spectrum of a pristine 2D AuNP array, red dots are the two excitation wavelengths  b) Transmission spectra of the AuNP array after irradiation on the interband transition c) Transmission spectra of the AuNP array after irradiation on the LSPR excitation. Reprinted and adapted with permission from ~\cite{Magnozzi19}. Copyright (2019) American Chemical Society.  }
	\label{Bisio_Melt}
\end{figure}

Another related experiment we have performed on these 2D arrays of gold NPs is to compare the efficiency of different types of excitation, either interband transition ($\lambda$ = 400 nm) or plasmonic excitation ($\lambda$ = 600 nm), in converting absorbed energy into thermal energy~\cite{Magnozzi19}.
These experiments were performed at much higher excitation intensities as the method to determine heating efficiency was
to examine permanent changes to the morphology and optical response of the NPs. At the high intensities used in this work the quantity of heat deposited
can lead to melting, coalescence, and even ablation of the NPs. Figure \ref{Bisio_Melt}a shows the 
transmission spectrum of the non-irradiated AuNP array where we mark in red the wavelengths of the incident laser radiation used in our experiment. At first 
glance one would expect the melting to be more efficient on plasmonic excitation due to the larger absorption at this wavelength (see 
transmittance measurements in figure \ref{Bisio_Melt}a) confirmed by comparison with theoretical modelling which found absorption cross sections of $\sigma_{abs}$(400 
nm) = 2.03 × 10$^{-16}$ m$^2$ and $\sigma_{abs}$(600 nm) = 3.45 × 10$^{-16}$ m$^2$, respectively, for these arrays~\cite{Magnozzi19}.

However, in the experiment exactly the opposite was found as can be seen from the microtransmission spectra of the areas irradiated by different fluences of 400 nm (figure \ref{Bisio_Melt}b) and 600 nm (figure \ref{Bisio_Melt}c) radiations: for the fluence of 75 J/m$^2$ the arrays are completely depleted by 400 nm radiation while only minor modifications of the transmission (and thus morphology of the AuNPs) occurs for the same fluence at 600 nm. 
These observation were confirmed by the AFM images of the irradiated areas~\cite{Magnozzi19}.
This unexpected behavior can be explained by taking into account the different distribution of the electric field in the NPs between the plasmonic and interband cases. Indeed, for interband excitation the electric field magnitude is relatively homogeneous, whereas for plasmonic interaction the electric field magnitude become strongly inhomogeneous with a strong concentration on the small portion of the NPs facing the interparticle gap (hot spots). For homogeneous fields, the NPs undergo homogeneous heating and consequently high temperatures are achieved throughout the NPs. For the plasmonic case, the total amount of energy deposited is larger due to the larger absorption cross section. A possible explanation is that highly localized heating due to the hot spots leads to ultrafast non-thermal melting~\cite{Harutyunyan15} and to the dissipation of the energy following different pathways such as local ablation of AuNP or the formation of plasma in hot spot volume due to photoionization, impact ionization and/or strong-field photoemission~\cite{Boulais12,Hobbs17}. These phenomena do not lead to efficient melting of the NPs privileging only localized reshaping of them. This demonstrates that interband excitation is significantly more efficient than plasmonic excitation in melting, coalescing, and sublimating the NPs.

In the context of the present article these results are encouraging as they suggest that processes which compete with simple heating of the NPs can occur in the case of plasmonic excitation. Further research is required to identify and, eventually, optimize these processes.

\begin{figure}[htb]
	\captionsetup{singlelinecheck = false, font=footnotesize, labelsep=space}
	\centering
	\includegraphics[scale=0.75]{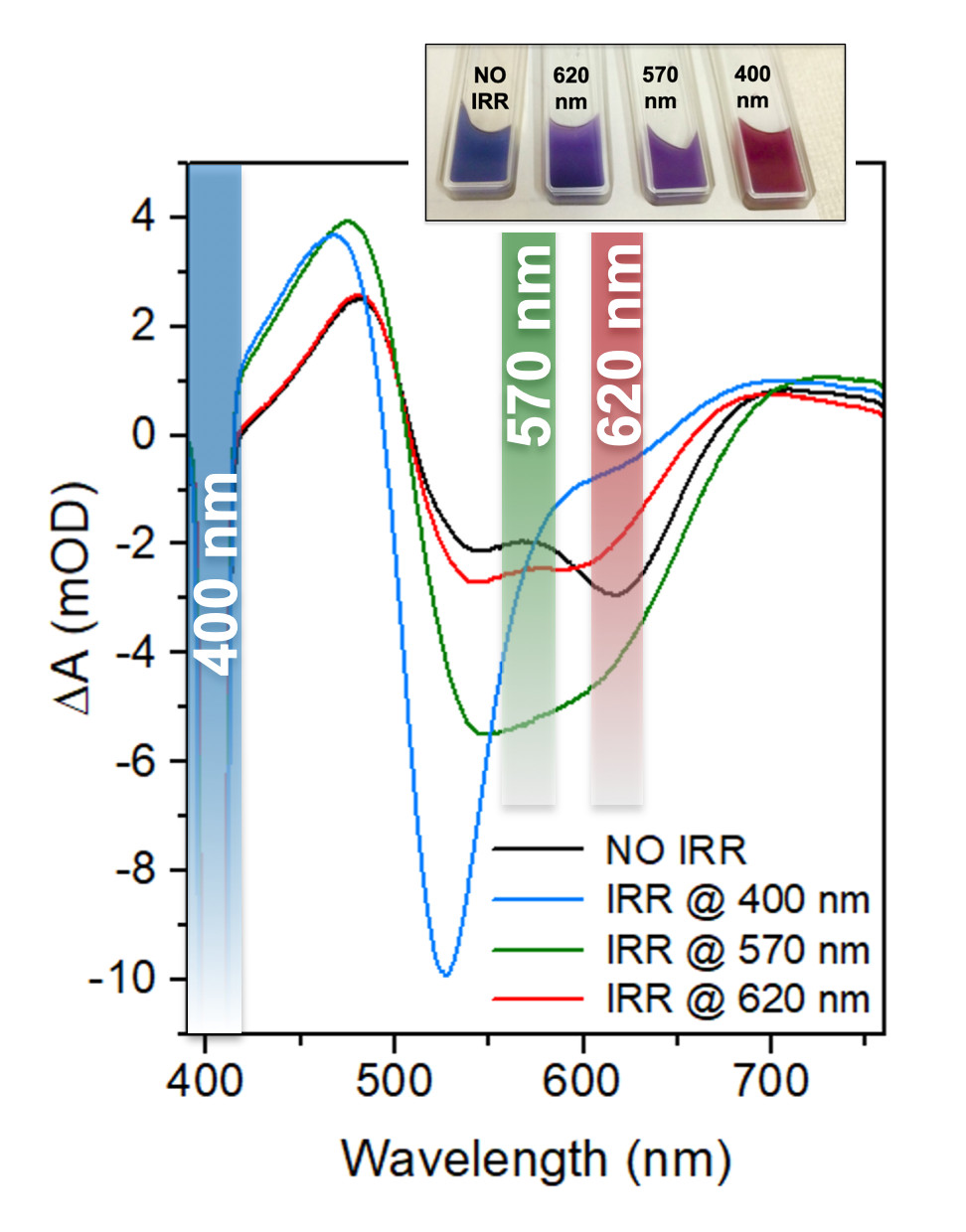}
	\caption{Effect of irradiance of the 10 nm gold spherical NPs, functionalized by rhodamine B 
		isothiocyanate samples at different wavelengths on the transient absorption at a delay of 1 ps.
		 The inset shows a photograph of the non-irradiated 
		and irradiated samples thus illustrating the effect of irradiation on the steady state absorption. }
	\label{NPs_sol2}
\end{figure}

For the final experiment we describe in this section we return to the 10 nm Au@RITC water solution described in the previous section. The reason for this is that the LSPR in these systems is broadened and shifted
to the red (see figure \ref{NPs_sol}b), with longer chains of NPs associated with higher red shifts. Therefore, by tuning the wavelength
of the exciting light it should be possible to exert control over the spatial distribution of energy deposited.
In this experiment we avoided too high laser fluences which may cause nanocavitation, explosive boiling of the water around the NPs or even fragmentation of the NPs~\cite{Hu04}
and chose fluences such that only small modifcations of the morphology of the aggregates occur. The induced modifications were subsequently investigated by measuring the new optical response and by SEM microscopy with the aim of understanding how the spatial distribution of the thermal energy deposited varies with excitation wavelength. Figure \ref{NPs_sol2} shows the TA of the non-irradiated Au@RITC water solution (black line), acquired at a pump-probe delay of 1 ps~\cite{Catone18}. The other colored lines show 
the TA acquired after irradiation of the solution at different wavelengths, corresponding to selective excitation in the 
interband (400 nm) or extended LSPR (570 and 620 nm) region. The macroscopic effect of this selective excitation can be clearly observed 
looking at the colors of the not-irradiated or irradiated solutions obtained (see inset of Figure \ref{NPs_sol2}).

Through finite integral technique calculations~\cite{Weiland77} we have shown that the spatial distribution and extent of the light concentration is strongly 
wavelength dependent: while interband irradiation causes a more isotropic and homogeneous heating leading to the coalescence into larger 
spherical NPs, plasmonic irradiation leads to a strong localization of the electric field in the interparticle regions, with hot spots playing 
a key role in promoting melting effects~\cite{Catone18}. Accordingly, by tuning the melting laser in the extended LSPR region, it was possible to 
selectively excite the different aggregate substructures present in the solution and thus choose the spatial distribution of the energy deposited. 

This experiment, which shows that it is possible to spatially localize the deposition of energy within plasmonic nanostructures, is important to keep in mind when designing/fabricating hybrid plasmonic/semiconductor nanostructures aimed at maximising
energy transfer from the plasmonic to the semiconductor material. Indeed, this is one of the reasons why embedding the plasmonic material in the semiconductor (see the AgNP@Ceria example below) may lead to high yield of energy transfer.

\subsection{Ultrafast Optical Response of Nanostructured Semiconducting Materials}

\subsubsection{Nanostructured Halide Organic-Inorganic Perovskites \\ }  

\begin{figure}[htb]
	 \captionsetup{singlelinecheck = false, font=footnotesize, labelsep=space}
	\centering
	\includegraphics[scale=0.9]{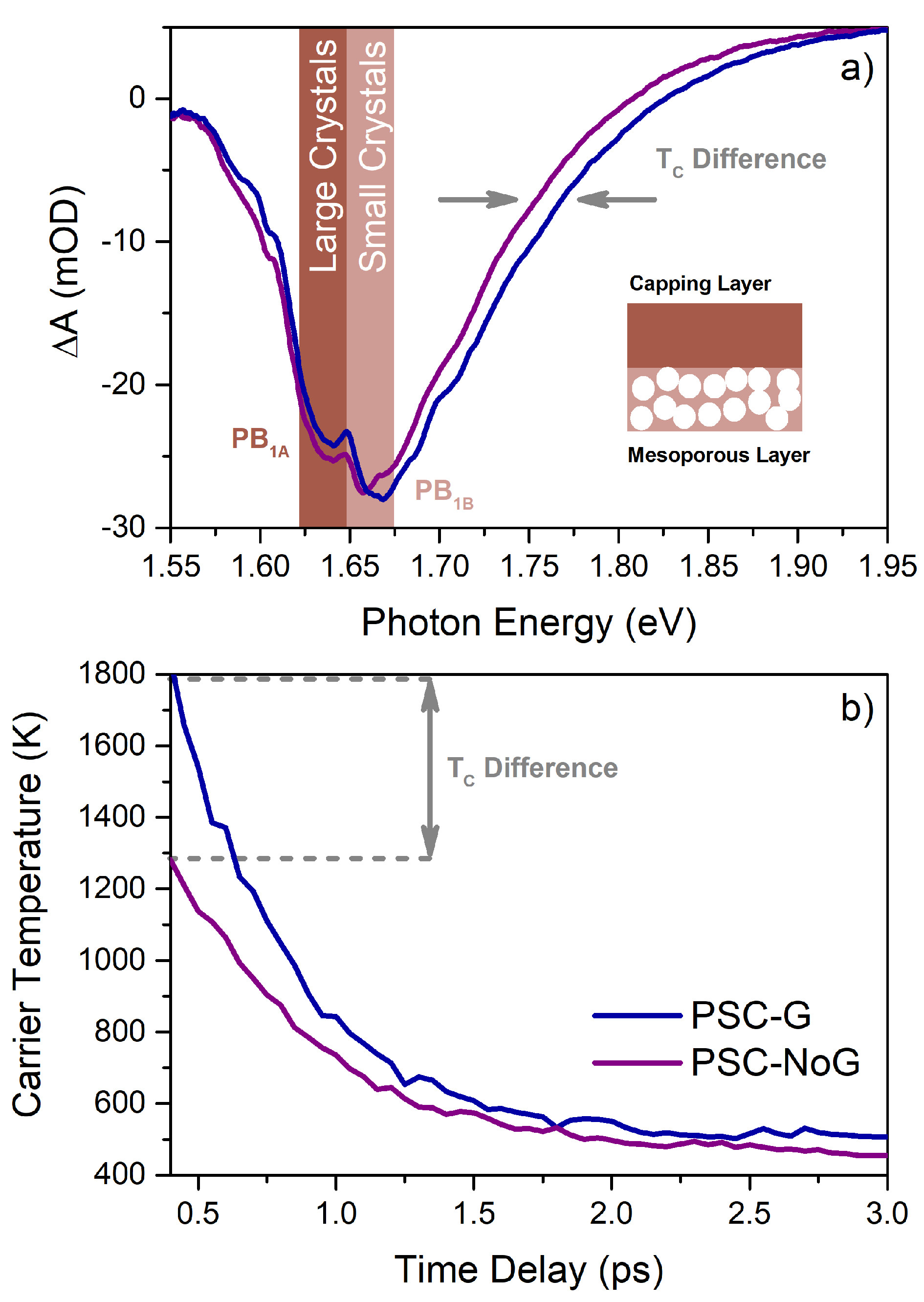}
	\caption{a) Transient absorption spectra acquired at 0.75 ps of time delay for 1-week aged PSC-G (blue line) and PSC-NoG (purple line). The photobleaching signal shows two distinct peaks at 1.64 eV (PB$_{1A}$) and 1.66 eV (PB$_{1B}$) assigned to the large crystals of the capping layer and the small crystals of the mesoporous layer, respectively. b) Carrier temperature as a function of time delay (from 0.4 to 3 ps) for 1-week aged PSC-G (blue line) and PSC-NoG (purple line). The cells were pumped at 3.1 eV with an initial carrier density of 7.0 x 10$^{17}$cm$^{-3}$. }
	\label{Perov}
\end{figure}

The morphology of polycrystalline films of semiconducting materials can have an important impact on the photoexcited carrier
dynamics of the materials. An example where this effect can have consequences for the functioning of a semiconductor based device is
in halide perovskite based solar cells (PSCs). The active layer of these devices is generally formed by spin coating precursor molecules
of the perovskite on to a substrate and allowing a crystallization process to form the perovskite layer. A major area of research is
the investigation of the effect of the morphology of this layer on device performance and stability~\cite{Zuo15,Fakharuddin17,Saliba16,Asghar17}. In the case that we investigated~\cite{Okeeffe19} the precursor molecules were 
coated onto a mesoporous TiO$_2$ (m-TiO$_2$) layer that acts as electron transport layer in the device.      
The result is a bilayer of perovskite consisting of a capping layer with large crystals (of $\sim$500 nm) and mesoporous layer with 
small crystals (20-40 nm) spatially restrained within the TiO$_2$ mesoporous layer (see inset of figure \ref{Perov}a). 
It is possible to distinguish the transient optical responses of these layers as they have slightly different absorption edges~\cite{DInnocenzo14} which leads to differences in their transient bleaching. The result is that photoexcitation of the large crystals leads to a bleaching signal centered at 1.64 eV (PB$_{1A}$) while excitation of the small crystals produces a bleaching at 1.66 eV  
(PB$_{1B}$) (see figure \ref{Perov} a).

The aim of our work was to study the effects of the addition of graphene flakes to the m-TiO$_2$ layer on the stability and carrier dynamics  of these perovsite layers. To do so we measured the transient absorption of two PSCs with (PSC-G) and without (PSC-NoG) graphene, on fresh 
samples and after one week of shelf ageing. While the relative intensities of PB1A and PB1B are unchanged in the spectra of PSC-G with respect to the 
as-prepared sample (data not shown~\cite{Okeeffe19}), they are different in PSC-NoG, suggesting that in PSC-NoG, a higher degradation of 
small crystals with respect to large crystals over time induces the intensity reduction of the PB$_{1B}$ with respect to PB$_{1A}$. 

Furthermore, the TA 
spectra presents another difference: the PB signal of PSC-NoG is less broadened (see gray arrows in figure
\ref{Perov}a) with respect to that of the PSC-G, which 
remains unchanged after aging. This high-energy tail is directly related to the carrier temperature (T$_C$) at a given time 
delay~\cite{Yang16,Yang17}. In fact, it is possible to extract the T$_C$ from the TA spectra by simply fitting the high-energy tail with a 
Maxwell-Boltzman function. This has been done and the T$_C$ as a function of time for the 1-week aged PSC-G and PSC-NoG is shown in figure \ref{Perov}b. It can be seen that the initial T$_C$ in the PSC-NoG (about 
1300 K) is significantly lower than that in PSC-G (about 1800 K). This behavior indicates that the T$_C$ difference arises from an overall faster thermalization of the 
carriers after excitation, occurring at time delays shorter than 0.4 ps. The conclusion is that the addition of graphene to the mesoporous layer protects the integrity of the 
small crystals, which in turn preserves the temperature of the carriers for a longer time after excitation.

\subsubsection{ZnSe and Si Nanowires \\} 

Among semiconductor nanostructures, nanowires are of particular interest for the study of semiconductor-metal interactions because 
the large aspect ratio of these nanostructures offers a large surface for the deposition of metallic nanoparticles and a small diameter 
enabling the interaction to be spread over a large portion of the semiconductor making any ensuing effect more easily observable than in bulk 
or thin films~\cite{Convertino14}. NW arrays favor light trapping making light absorption 
more efficient because of the reduced reflectivity~\cite{Street08,Street09,Convertino10,Convertino12}. This latter feature, along with the use 
of opaque substrates for the NW growth, make it difficult to perform transient reflectance or absorption measurements using conventional NW 
growth methods. In order to allow for transient absorption measurements, we have used NWs grown on transparent substrates. As these kinds of 
substrates do not withstand high temperatures, we turned our attention to materials that can be grown at low temperature with good optical 
quality, as ZnSe~\cite{Zannier14} and Si NWs~\cite{Tian16b}. This type of samples are semitransparent and allow for a good signal-to-noise 
ratio in the transient measurements~\cite{Tian16,Tian19}.

\begin{figure}[htb]
	\captionsetup{singlelinecheck = false, font=footnotesize, labelsep=space}
	\centering
	\includegraphics[scale=1.1]{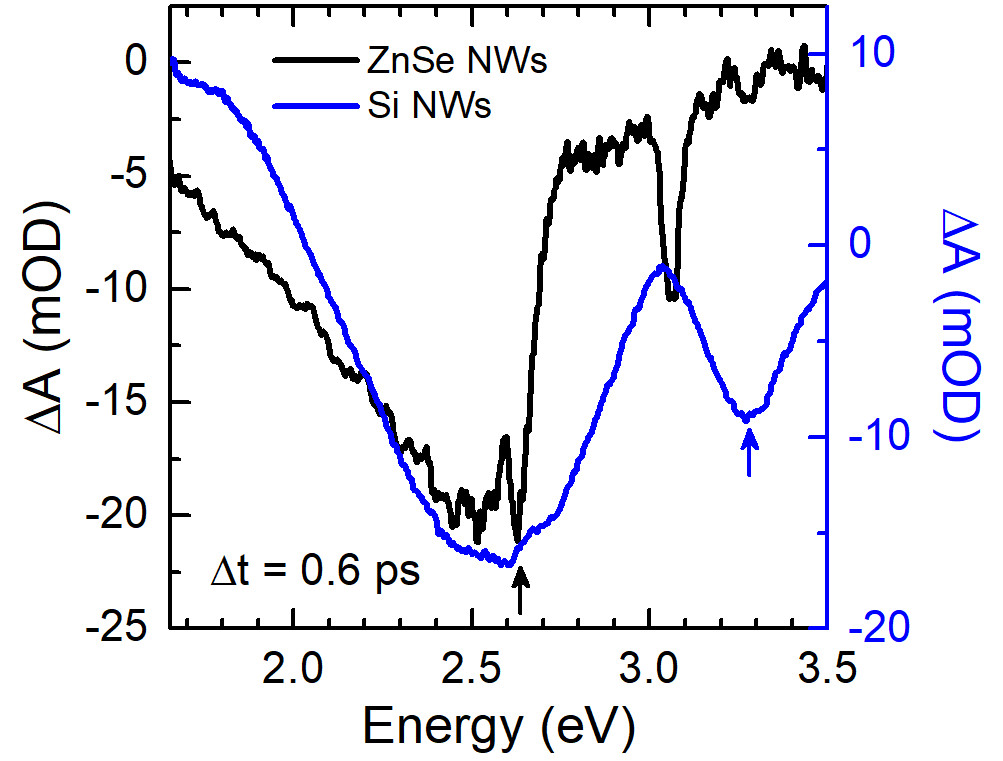}
	\caption{a) Transient absorption spectra acquired at a time delay of 0.6 ps by photoexciting a sample of ZnSe nanowires at 3.05 eV (black line) and a sample of Si nanowires (blue line) at 4.5 eV. The arrows indicate the energies of band edge bleaching. The sharp peak at 3.05 eV in the ZnSe nanowire spectrum is an artifact due to scattering of the pump radiation. }
	\label{NWs}
\end{figure}

As mentioned in the introduction in time-resolved optical measurements of semiconductors the most typical spectral feature consists of a absorption bleaching at energies 
corresponding to excitonic or band gap transitions. This negative signal is due to decreased absorption at those energies induced by the photoexcited 
carrier population. This is clearly observed in both ZnSe and Si NWs, as shown in figure \ref{NWs}, with some 
important differences between the two materials. ZnSe is a direct gap material with a band gap of 2.6 eV at room temperature and we observe 
indeed a sharp bleaching signal at this energy (see the black arrow in figure \ref{NWs}), that is hence attributed to near band-edge (NBE) transitions. A broad bleaching signal is also 
observed at lower energies due changes in the absorption given by point defects, as usually also observed in epitaxial ZnSe, that are reduced 
at low growth temperatures~\cite{Imai88}. Si is an indirect gap material with energy of 1.1 eV, at which a clear absorption bleaching is not 
observed~\cite{Tian19} while a strong and sharp bleaching is observed at 3.3 eV  (see the blue arrow in figure \ref{NWs}), energy corresponding to the direct band gap of Si. The strong 
negative signal observed for Si below the direct band gap energy is due to the modification of the whole Si band structure upon 
photoexcitation~\cite{Sangalli16}. Another difference between these two materials is the rise time of the bleaching signal. While for ZnSe we 
measure 170+/-60 fs~\cite{Tian16}, we are not able to measure it for Si as it is faster than the IRF of our set up. This latter finding is in 
agreement with what found by Schulze and coworkers, who measured the band gap shrinking occurring in hundreds attoseconds~\cite{Schultze14}.

\subsection{Energy Transfer in Hybrid Nanostructured Plamonic/Semiconducting Materials}

\subsubsection{3D Arrays of Gold and Silver Nanoparticles deposited on ZnSe Nanowires \\ }

\begin{figure}[htb]
	\captionsetup{singlelinecheck = false, font=footnotesize, labelsep=space}
	\centering
	\includegraphics[scale=0.7]{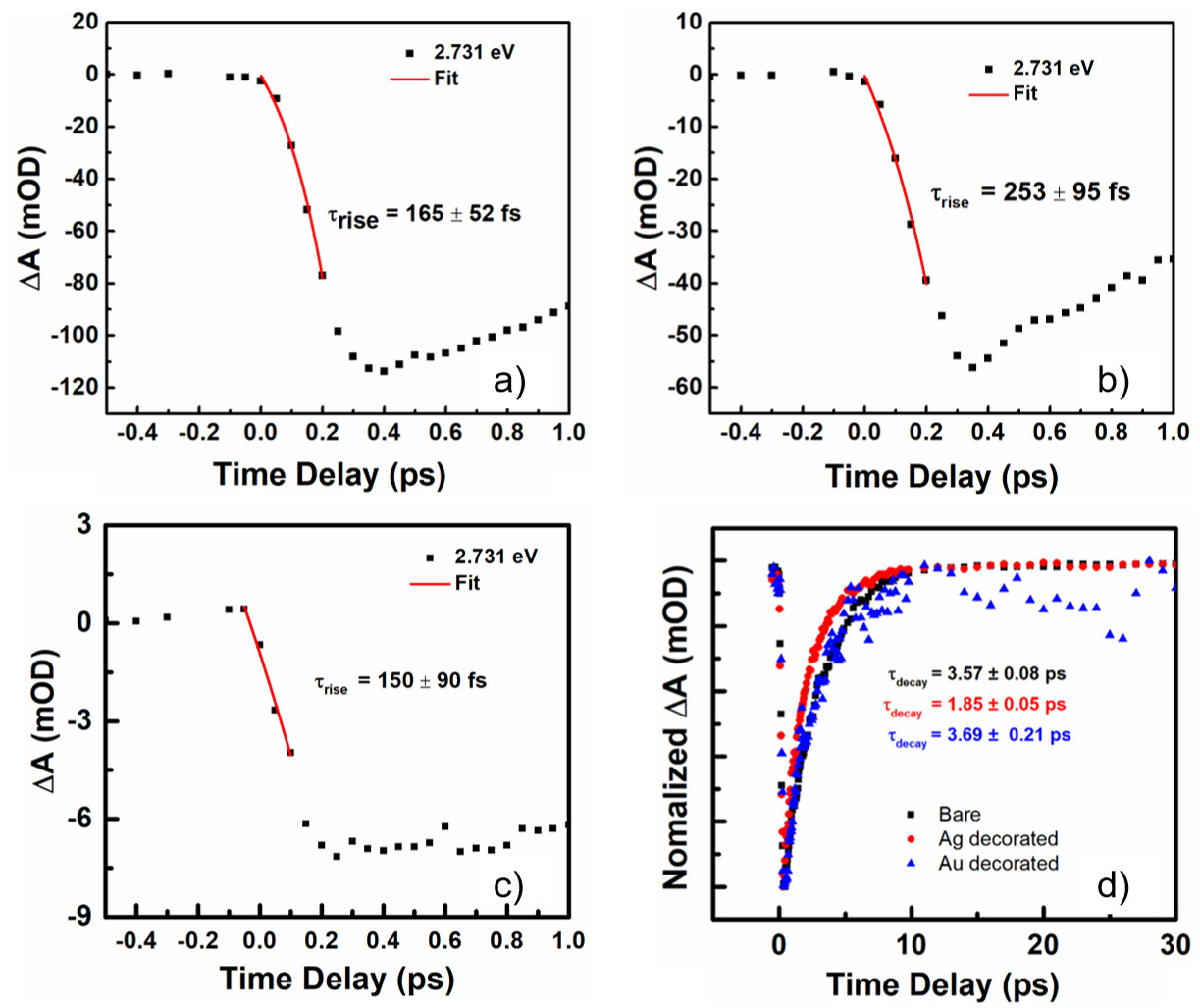}
	\caption{ a) Rise time of near band-edge bleaching in bare ZnSe nanowires (ZnSe NWs), b) Rise time of near band-edge bleaching in Ag NP decorated ZnSe nanowires (AgNP@ZnSeNWs), c) Rise time of near band-edge bleaching in Au NP decorated ZnSe nanowires (AgNP@ZnSeNWs), and d) Comparison of the decay times of
	band edge bleaching in bare ZnSe NWs, AgNP@ZnSeNWs and AuNP@ZnSeNWs. These plots show how the AgNPs modify the dynamics of the ZnSe NW while the AuNPs do not. }
	\label{AgNP_ZnSeNW}
\end{figure}

The dynamics of the absorption bleaching in ZnSe can be varied by the presence of Ag NPs on the NW sidewalls. Both rise and decay times of the bleaching signal are affected by the presence of the Ag NPs. In particular, as shown in figure \ref{AgNP_ZnSeNW}, the rise time increases from about 170 fs (see figure \ref{AgNP_ZnSeNW}a) to about 250 fs (see figure \ref{AgNP_ZnSeNW}b), while the decay time halves decreasing from 3.5 ps to 1.85 ps (see figure \ref{AgNP_ZnSeNW}d). We notice here that no change in the dynamics of the bleaching signal is instead observed if Au NPs are formed on the ZnSe NW sidewalls (see figures \ref{AgNP_ZnSeNW}c and \ref{AgNP_ZnSeNW}d). As discussed above, electronic interaction between a semiconductor and a plasmonic NP mainly occurs through a number of pathways that depend upon the spatial proximity and the spectral overlap between the LSPR of the semiconductor transitions. The mechanisms based on direct and indirect hot-carrier injection and FRET~\cite{Forster65,Li15}, benefit from the proximity of the materials. However, for FRET, a necessary condition is the spectral overlap between the LSPR and the absorption band of the semiconductor ~\cite{Forster65}. In Ag-decorated ZnSe NWs, there are spectral overlaps and physical contact between the metal and the NW. This makes possible both FRET and hot-carrier transfer. In Au-decorated ZnSe NWs, where we have observed no change in the bleaching dynamics, there is contact between metal and NWs, but no spectral overlap between Au LSPR and ZnSe NBE absorption.

The decay of the NP LSPR produces hot electrons in both Ag and Au NPs, whose energy distribution will reflect the density of states in the two metals. Because of the band alignment between the metals and ZnSe~\cite{Vos89,Pelucchi06}, there are very close similarities between the Ag/ZnSe NW and Au/ZnSe NW interfaces and only part of the hot electrons will have energy above the ZnSe conduction band. Therefore, in the case the effects we observe in Ag-decorated ZnSe NWs were due to hot-carrier transfer, we should observe the same feature in Au-decorated ZnSe NWs but this is not the case. Hence, a resonant mechanism must underlie the modifications observed with the formation of the Ag NPs on the NW sidewalls. In particular, the increase in the rise time of the bleaching signal points toward the presence of a FRET mechanism, where the Ag NPs act as donors and the ZnSe NWs are the acceptors~\cite{Forster65,Li15}. The increase of the rise time can be due to the excitation of electron-hole pairs in ZnSe at a later time than for direct photoexcitation, because of the energy transfer from the NPs toward the NWs. 

Instead, a decrease of the decay time of the absorption bleaching is observed over the whole measured energy range. The absorption spectrum of the Ag NPs is very broad, due to the presence of both dipole and quadrupole excitation of the plasmon and, as shown above, of the shape distribution of the NPs~\cite{Dimario18,Kelly03,Liu15}. At low energies, free-carrier absorption also contributes to their absorbance. The broad width may allow the resonant interaction with the defect band, with effects similar to those observed for the NBE. This resonant interaction with the defects, via an increased carrier-phonon scattering rate would explain the faster decay time in the TA signal of the Ag- decorated samples.

\subsubsection{Silver Nanoparticles Embedded in CeO$_2$ Thin Films \\ }

The study of photocatalysis mediated by the LSPR has generated enormous interest in recent years due to the opportunity of extending the energy 
range of activity of traditional semiconductor photocatalysts when coupled with metallic nanoparticles~\cite{Clavero14,Zhang18,Kale14,Narang16}. Among the semiconductor 
catalysts, cerium oxide (CeO$_2$) is attracting growing interest due to its efficiency in catalyzing redox reactions~\cite{Navalon11,Lei15} thanks to the presence of localized Ce 4f states between the O 2p valence band and the Ce 5d conduction band~\cite{Skorodumova02}. In particular, we have used TA spectroscopy to investigate the mechanisms of LSPR decay in silver NPs embedded in a thin film of CeO$_2$ (Ag@CeO$_2$).

\begin{figure}[htb]
	\captionsetup{singlelinecheck = false, font=footnotesize, labelsep=space}
	\centering
	\includegraphics[scale=0.65]{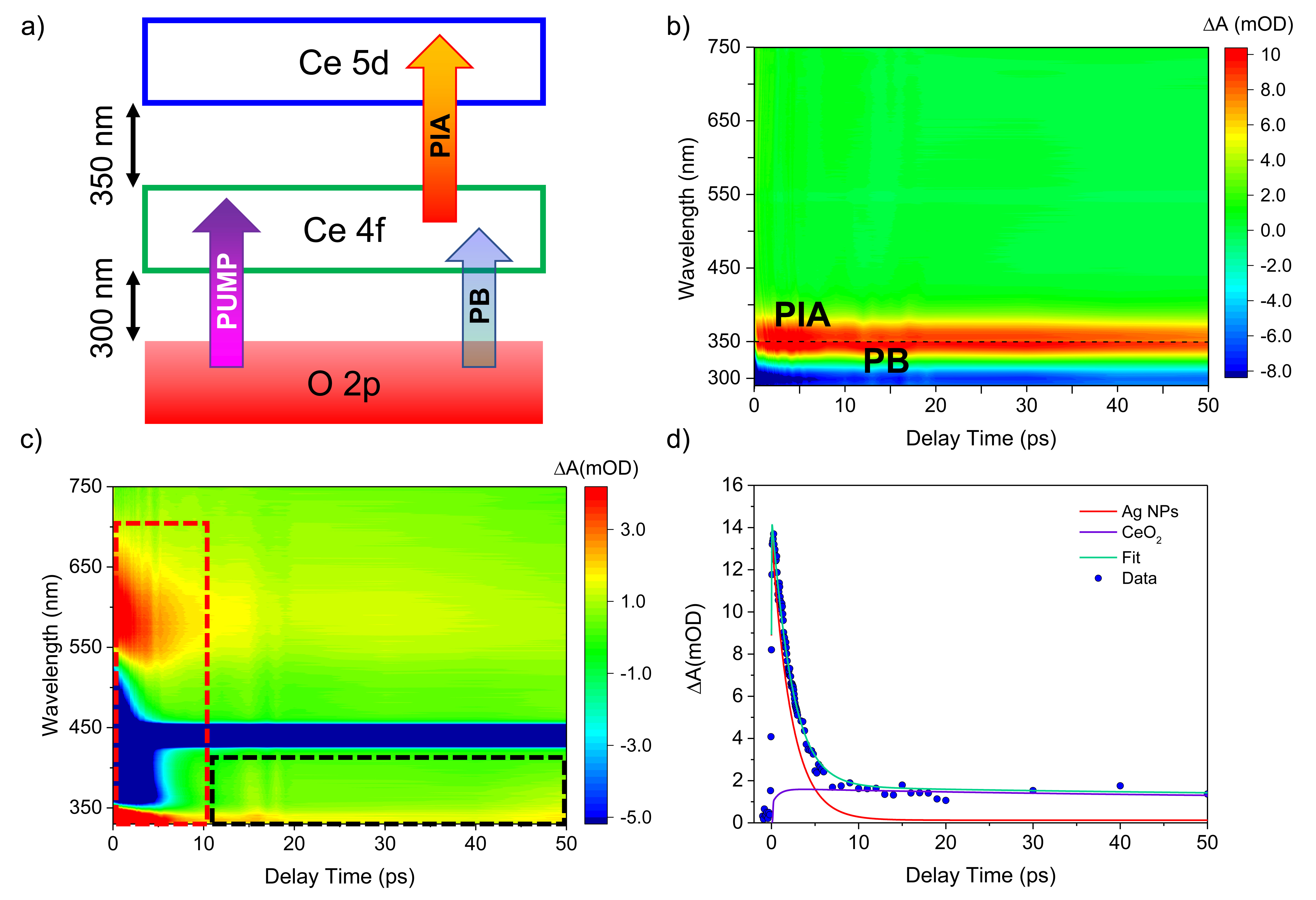}
	\caption{ a) Sketch of the CeO$_2$ electronic structure, showing the transitions induced by the excitation of the pump at 275 nm and the corresponding TA signals (PB and PIA). b) False-color map of the TA spectra for the CeO$_2$ sample excited with a pump at 275 nm. The black dashed line at 350 nm separates the two sets of data acquired with the UV and Visible white light supercontinuum probes. c) False-color maps of the TA spectra of the Ag@CeO$_2$ sample excited with a pump at 440 nm, below the CeO$_2$ band gap and at LSPR maximum. The red dashed box indicates the part of the map dominated by the plasmonic response of the Ag NPs while the black dashed box delimits the PIA signal due to the electron injection into the CeO$_2$ film. The negative signal at 440 nm is due to scattering of the pump. d) Experimental time dependence of the transient PIA signal at 345 nm of the Ag@CeO$_2$ sample pumped at 440 nm. The components used for data fitting are shown as solid lines: PIA of CeO$_2$ (purple line); the plasmonic deexcitation (red line); the full fitting spectrum (green line). }
	\label{AgCeria}
\end{figure}

Figure \ref{AgCeria}b shows the false-color map of TA spectra of the CeO$_2$ film without Ag NPs, excited with a pump at 275 nm (4.5 eV), above the optical gap of the semiconductor. The TA map is dominated by two features: a negative peak at 300 nm and a positive peak at 345 nm, showing only 
a slight decrease in intensity in the temporal window explored. This behavior suggests that the excited electrons are trapped in localized 
states~\cite{PelliCresi20}. The negative peak is assigned to the bleaching of the band edge, considering the band structure of CeO$_2$, as 
schematically reported in Figure \ref{AgCeria}a. With this consideration in mind, we have assigned the positive peak to photoinduced absorption (PIA) from the Ce 4f levels into the 
Ce 5d band by the probe (see arrows in Figure \ref{AgCeria}a). When we excited Ag@CeO$_2$ with a pump at 275 nm, above the optical gap of CeO$_2$ and the 
inter-band transition of Ag NPs, we obtained a TA map with features related not only to the excited ceria but also to the transient 
plasmonic response of Ag NPs (data not shown~\cite{PelliCresi19}). In particular, the TA map presents a broad positive band around 610 nm and a broad negative 
band at 380–520 nm, associated with the transient line shape of the excited LSPR that presents the typical plasmonic decay dynamics in the first 10 ps similar to the TA response of the AuNPs in solution described above. 
These features are still clearly visible in the TA map also when we have excited the sample with a pump at 440 nm, resonant with the LSPR and 
below the optical gap of CeO$_2$, highlighted by the black dashed box in figure \ref{AgCeria}c. Moreover, even if the energy of the pump is not enough to excite the 
CeO$_2$ above the optical gap, the TA map of Ag@CeO$_2$ exhibits at time delays higher than 10 ps the same persistent signal at 345 nm assigned to PIA 
in CeO$_2$ (see red dashed box in Figure \ref{AgCeria}c). The presence of this signal has been assigned to transient occupation of 4f levels in ceria and is induced by LSPR-mediated electron injection.

The quantification of the efficiency of the electron injection was derived by comparing the PIA signal acquired at different pump energies. In 
particular, the temporal cuts of the TA maps at 345 nm were fitted by a linear combination of a contribution of the bare ceria, obtained from 
the thin film of CeO$_2$~\cite{PelliCresi19}, and an exponentially decreasing function related to the positive signal generated by the excitation of the LSPR, as 
reported in Figure \ref{AgCeria}c for Ag@CeO$_2$ pumped at 440 nm. The signal related to the plasmonic dynamics dominates at delay times shorter than 5 ps, 
preventing the possibility to access the dynamics of the electron injection in this temporal region. For this reason, the quantification of 
the injection efficiency was obtained thanks to the TA signals at 345 nm of the Ag@CeO$_2$ sample pumped at different energies below the optical gap 
in the 50–250 ps delay time range, where the LSPR-related signals are negligible~\cite{PelliCresi19}. The resulting injection efficiency values are in the 
range of 12–16\% for excitation between 400 and 500 nm, while a lower value of 6\% is found for the pump at 600 nm~\cite{PelliCresi19}. As already discussed 
in the introduction, there are a number of mechanisms that can induce an electron injection mediated by the excitation of the LSPR, however in contrast to the previous example of AgNP@ZnSe NWs, the mechanisms that require an overlap between the LSPR and an optical 
transition in the semiconductor (light trapping, energy transfer to the semiconductor through near-field enhancement, plasmon induced 
radiative energy transfer and resonant energy transfer~\cite{Atwater10}) can excluded because they cannot significantly contribute in this case. 
For this reason, the only mechanisms that are taken into consideration are the plasmon-induced indirect hot electron injection~\cite{Clavero14} and 
plasmon-induced direct electron injection~\cite{Wu15,Long14}. The efficiency of the indirect process is intrinsically limited by competition with 
electron thermalization that leads to a hot Fermi–Dirac distribution in which most electrons do not have
sufficient energy to overcome the Schottky barrier, and by the fact that the electrons need the moment vector directed towards the interface 
to be injected into the semiconductor. The combination of these two limitations means that, even in the most optimistic of estimates, a maximum of 8\% injection efficiency can be achieved~\cite{White12}. In contrast, the plasmon-induced direct carrier injection has been predicted to have efficiencies of between 40 and 50 \%~\cite{Long14,Ma19} in TiO$_2$ sensitized with metal NPs. Furthermore the mechanism has been experimentally confirmed~\cite{Wu15,Tan17} as a mechanism with a very high 
efficiency. For these reasons, we suggest that the high efficiencies of electron
injection observed in the Ag@CeO$_2$ system are at least partly due to this direct mechanism~\cite{PelliCresi19}.
Of course experimentally we cannot exclude the participation of the indirect mechanism and, indeed, to distinguish between these 
mechanisms would require experiments with resolution on the sub 10 fs timescale.

\subsection{Alternative Method for Sensitization of Wide Band-Gap Semiconductors to Visible Light  }

\subsubsection{V-doped TiO$_2$ Nanoparticles \\ }

\begin{figure}[htb]
	\captionsetup{singlelinecheck = false, font=footnotesize, labelsep=space}
	\centering
	\includegraphics[scale=0.45]{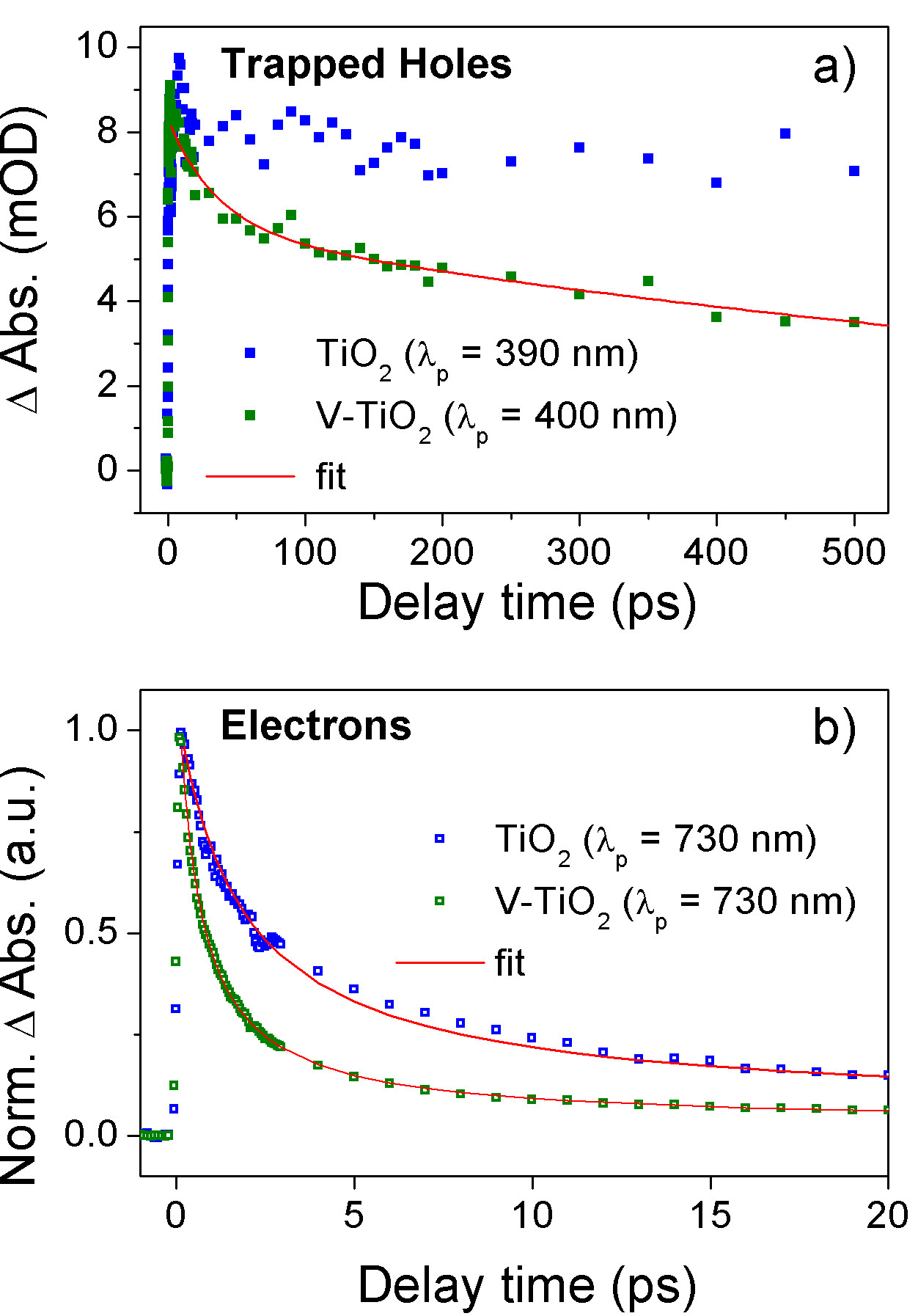}
	\caption{Observed dynamics following excitation of TiO$_2$ and V-TiO$_2$ NPs at a wavelength of 330 nm. a) dynamics of trapped holes, b) dynamics of photogenerated electrons. Both panels show the increased rate of recombination
		in the V-TiO$_2$ NPs in comparison to the TiO$_2$ NPs.  }
	\label{VTiO2}
\end{figure}

Ultrafast optical spectroscopy can also be used to investigate alternative methods to sensitize wide band-gap semiconductor materials to visible light which do not involve plasmonic materials. For example, a common technique to extend the absorption 
of TiO$_2$ into the visible range is to dope the material with
a range of elements~\cite{Pelaez12}. An example of this is our study of the
charge carrier dynamics in undoped and vanadium-doped TiO$_2$ nanoparticles (TiO$_2$ and V-TiO$_2$ NPs, respectively)~\cite{Rossi18}. 
The aim of this work was to understand the link
between the ultrafast charge carrier dynamics and the photocatalytic efficiency of the materials. 
To understand the effects of the doping on the carrier dynamics we recorded 
TA spectra of the NPs deposited on a quartz substrate in the 360-770 nm range following 
ultra bandgap excitation (330 nm) and near and sub bandgap excitation (400 nm and 520 nm). The transient 
spectra of TiO$_2$ have been extensively studied in the past~\cite{Yoshihara04,Tamaki07} and these studies guided our assignment of the
observed spectral features of the TA: the peak at 400 nm was assigned to absorption of trapped holes,
a monotonically increasing absorption towards the IR was assigned to free electron absorption and a    
TA band at 730 nm was assigned to absorption of trapped electrons. The latter signals were impossible to separate in
our measurements and therefore we refer to the absorption at 730 nm as being due to both trapped and free electrons.
The dynamics observed following ultraband gap excitation (see Figure \ref{VTiO2}) reveal a more rapid decay of the
transient signals assigned both to trapped holes and to electrons in the V-TiO$_2$ NPs suggesting that the recombination
mechanisms are promoted by the presence of the V defects. Near band gap excitation showed five times higher signals in the V-TiO$_2$ 
NPs samples for excited carriers while only the V-TiO$_2$ sample showed excited carrier signals for the sub band gap
excitation.

These observations were correlated with the measurements of the internal photon conversion efficiency (IPCE) for photoelectrochemical
water splitting of electrodes made with these materials. The IPCE of the TiO$_2$ NP based photoanodes is higher than the V-TiO$_2$ NP based photoanodes for photoexcitation wavelengths $<$ 350 nm while the reverse is true for photoexcitaion $>$ 350 nm. This suggests that
the increased recombination in the V-TiO$_2$ NPs following ultrband gap excitation leads to a decrease in the photoelectrocatalytic
efficiency as the number of carriers available for photocatalysis is reduced while in the sub band gap excitation the increased absorption
leads to a higher efficiency for the V-TiO$_2$ NPs. This demonstrates the contribution of the optical measurements of the carrier dynamics to the understanding the functional efficiency of materials.    

\section{Conclusion and Perspectives}

The optical pump-probe measurements described in this article illustrate the kinds of information
on the photoexcited electron/carrier dynamics that can be accessed for nanostructured semiconductor
and plasmonic materials using this technique. In the case of plasmonic gold NPs we have discussed how to follow the ultrafast relaxation 
processes which occur following plasmonic excitation by using the time dependent optical response
of the materials. While in the case of the semiconductors we have discussed the band edge 
bleaching dynamics of ZnSe and Si NWs, and the morphological effects on hot-carrier cooling (hybrid organic 
inorganic perovskite thin film).
Ideally what is required in a hybrid system, in which the absorption cross-section of the
plasmonic material is used to sensitize the semiconductor to the exciting light, is that energy is transferred from
the plasmonic material before the energy is lost to these ultrafast processes and is ultimately dissipated 
as heat. In this work we discussed two examples in which such ultrafast energy transfer occurs: i) Ag NPs on ZnSe
NWs in which the spectral overlap of the plasmon resonance and the band gap of the semiconductor allows
energy transfer to occur via the FRET mechanism and ii) Ag NPs embedded in a CeO$_2$ thin film in which we detected
a highly efficient charge transfer following plasmonic excitation of the Ag NPs to long-lived states in the 
CeO$_2$ film. Finally, an alternative method of extending the absorption of a wide band semiconductor into the visible 
range is discussed based on the doping of TiO$_2$ NPs with vanadium and the subsequent effect on 
carrier dynamics and photoelectrocatalytic efficiency.

As can be seen from the above examples of hybrid plasmonic/semiconductor materials the energy transfer occurs on an ultrafast timescale and 
clearly ultrafast techniques such as the one described here are required to study such processes.
On the other hand, the desired photoinduced functional properties of the materials often take place under continous wave excitation (solar cells, catalysis) and therefore it can be difficult to directly relate the ultrafast processes to the efficiencies of devices based on these materials. For example, catalysis can take place on the microsecond to second timescale
and therefore in future experiments it will be important to bridge the temporal gap
by correlating the ultrafast processes with improved performance of the material on the long timescales
required for improved functionality. An example of this kind of approach can be found in our study of
the TiO$_2$ and V-TiO$_2$ described above, however, this approach is not widely employed in the literature.   
Indeed, this has been identified as one of the reasons for the lack of technological advance in
photocatalytic water treatment in spite of great deal of ``initial academic hype" on high energy transfer
efficiencies~\cite{Loeb19}.  

Another interesting area of research which is little explored in the literature is the investigation
of the effects of externally applied fields on interface transfer dynamics. External fields
can modify band alignment and barriers at the interfaces thus strongly modifying charge transfer 
processes. Such modified transfer efficiencies will certainly be more relevant to the operational
efficiency of devices which involve the external application of fields such a solar cells and
photoelectrocatalysis. 

Finally, other future research will be aimed at understanding the combined effect of 
plasmonically generated charge transfer and the thermal effects due to losses in the plasmonic material
on photocatalytic efficiency. In the recent literature there has been significant debate on
which of these mechanisms dominates in plasmonically assisted photocatalysis in metallic plasmon
materials~\cite{Dubi19,Dubi20,Gargiulo19}. Obtaining a synergy between the two mechanisms~\cite{Li20} 
may lead to higher overall efficiency the hybrid systems discussed in this work.

\section{Acknowledgements}
	
This work has received funding from the European Union’s 7th Framework Programme for research, technological development, and demonstration under Grant No. 316751 (NanoEmbrace) and from the Horizon 2020 program of the European Union for research and innovation, under grant agreement no. 722176 (INDEED). We thank Silvia Rubini and Valentina Zannier (IOM-CNR, Trieste) for providing the ZnSe NWs, Francesco Bisio (SPIN-CNR) for providing the 2D arrays of NPs, Federico Boscherini and Luca Pasquini (University of Bologna) for the supplying the TiO$_2$ NPs, Iole Venditti (Roma Tre University) and Ilaria Fratoddi (Roma Tre and "La Sapienza" University) for the Au NPs in solution, Paola Luches (NANO-CNR) and Sergio D’Addato (UniMore) for the Ag@Ceria films, Aldo DiCarlo, Antonio Agresti and Pescetelli (Tor Vergata University) for the perovskite samples, Bruno Palpant (LPQM) and Remo Proietti Zaccaria (IIT) for theoretical support and Davide Sangalli and Andrea Marini (ISM-CNR) for very fruitful discussions.

This is the version of the article before peer review or editing, as submitted by an author to Nanotechnology. IOP Publishing Ltd is not responsible for any errors or omissions in this version of the manuscript or any version derived from it. The Version of Record is available online at https://doi.org/10.1088/1361-6528/abb907 	

\section{References}

\bibliography{Focus57}
\bibliographystyle{iopart-num}

\end{document}